\documentclass[aps,a4paper,superscriptaddress]{revtex4}
\usepackage{graphicx}
\usepackage[all]{xy}
\usepackage{amsmath}
\usepackage{amssymb}

\newcommand{\be}{\begin{equation}}
\newcommand{\ee}{\end{equation}} 
\newcommand{\ben}{\begin{eqnarray}}
\newcommand{\een}{\end{eqnarray}}
\newcommand{\bes}{\begin{subequations}}
\newcommand{\ees}{\end{subequations}}

\newcommand{\LL}{{\cal L}}

\begin{document}

\title{Cosmology in the Universe with distance dependent Lorentz-violating bakground}
\author{C. A. G. Almeida}
\email{cviniro@gmail.com}
\affiliation{Departamento de Ci\^encias Exatas, Universidade Federal
da Para\'{\i}ba, 58297-000 Rio Tinto, PB, Brazil}

\author{M.A. Anacleto}
\email{anacleto@df.ufcg.edu.br}
\affiliation{Departamento de F\'{\i}sica, Universidade Federal
da Campina Grande, 58109-970 Campina Grande, PB, Brazil}

\author{F.A. Brito}
\email{fabrito@df.ufcg.edu.br}
\affiliation{Departamento de F\'{\i}sica, Universidade Federal
da Campina Grande, 58109-970 Campina Grande, PB, Brazil}
\affiliation{Departamento de F\' isica, Universidade Federal da Para\' iba,\\  Caixa Postal 5008, Jo\~ ao Pessoa, Para\' iba, Brazil}

\author{E. Passos }
\email{passos@df.ufcg.edu.br}
\affiliation{Departamento de F\'{\i}sica, Universidade Federal
da Campina Grande, 58109-970 Campina Grande, PB, Brazil}

\author{J.R.L. Santos}
\email{joaorafael@df.ufcg.edu.br}
\affiliation{Departamento de F\'{\i}sica, Universidade Federal
da Campina Grande, 58109-970 Campina Grande, PB, Brazil}

\begin{abstract}
We consider a cosmological setup with the inflaton field in the presence of a redshift dependent Lorentz-violating time-like background to address the inflationary regime and other phases of the Universe. We also show that the regime of dark energy at large distances (low redshifts) is essentially dominated by the presence of the Lorentz-violating background.
\end{abstract}
\maketitle

\section{Introduction}

One of the first main issues in cosmology was explaining the problem of the horizon, flatness and magnetic monopole, which was successfully explained with the advent of the inflationary phase in  the early Universe \cite{Guth:1980zm,Linde:1981mu,linde_14}. This accelerated phase should end with the reheating of the Universe coinciding with the beginning of a radiation-dominated era and then a (dark) matter-dominated to forming the first structures such as galaxies and clusters. These are decelerated phases. However, in the year 1998, Riess et al. \cite{Riess:1998cb} and Perlmutter et al. \cite{Perlmutter:1998np} independently reported a present acceleration of the Universe. The source of this acceleration was denominated {\it dark energy}. Since then, several models have been considered in the literature in order to accomplish this new observation into the standard cosmology. Although its origin is not known yet some aspects are similar to inflationary phase, as for example, its pressure should be negative in order to counteracting the gravitational force and then providing accelerated expansion. The simplest candidate for the dark energy is the cosmological constant with the equation of state $\omega=-1$.  However, it might well be possible that the origin of the dark energy is not due to a cosmological constant. If so it is necessary to look for alternative models to explain the present accelerated expansion of the Universe. Yet, as in the inflationary phase, one can model the source of the dark energy with scalar fields. The main difference is the scale too low compared with the inflation scale and that the scalar potentials should be flat enough to drive the present accelerated expansion of the Universe. The models with scalar field describing dark energy are basically those called quintessence \cite{q1,q2,q3,q4,q5,q6,q7,q8} and k-essence\cite{k1,k2,k3}. In the former case one focus on the use of scalar field with slowly varying potentials, while in the latter case is the kinetic part that drives the present accelerating phase of the Universe. One should mention also that these are not the only way to model the dark energy. To quote the main recent alternatives, we mention the models with modified gravity such as $f(R)$ gravity \cite{fr1,fr2,fr3}, $f(R,T)$ gravity \cite{harko_11,ms_16}, scalar-tensor theories \cite{st1,st2,st3,st4,st5} and braneworlds models \cite{brane1,brane2}. One might consider in this class of theories the $\Lambda$CDM model which is based on Einstein or $f(R)$ gravity with a cosmological constant. Nowadays all these models can be distinguished from each other through the use of observational data such as combined analysis of SN Ia, CMB and BAO \cite{Riess:1998cb,Perlmutter:1998np,WMAP,planck15}. These observations have constrained the equation of state of dark energy to the bound $-1.097<\omega<-0.858$ at $95\%$ of confidence \cite{WMAP,planck15}. This favors the cosmological constant as a good candidate of dark energy, whereas disfavors several other types of models. However, further observations confirming or detecting deviations of the $\Lambda$CDM model are very important to shed new light in the origin of the dark energy.

The study of several Lorentz-violating cosmological scenarios in the inflationary phase have been addressed in the literature \cite{gasp,Lim:2004js,Li:2007vz,Zuntz:2008zz,ArmendarizPicon:2010rs,Kanno:2006ty,Donnelly:2010cr,bazeia}. {Between such works we can highlight as motivations for our investigation, the approaches from Gasperini \cite{gasp}, and from Donnelly and Jacobson \cite{Donnelly:2010cr}, for instance. In his work Gasperini proposes that the primordial phase of accelerated expansion of the Universe could be achieved if at some very early epoch the gravitational interactions were described by a non locally Lorentz invariant theory. It is also suggested that this additional mechanism for producing inflation could be used to solve some problems of the standard inflationary scenario. Besides, in Ref.~\cite{Donnelly:2010cr} the authors consider a Lorentz-violating theory of inflation formed by an Einstein-aether theory coupled with a scalar field Lagrangian. There the authors determined cosmological parameters which are affected by the Lorentz violation but still allow for a natural end to inflation. On the other hand, application of such a scenario can also be found  in dark energy, where is shown that violation of the Lorentz invariance induces Lagrangians that are able to drive the current acceleration of the universe \cite{Blas:2011en,Audren:2013dwa}. As such, the former and the latter theories deal with Lorentz symmetry violation at small and large distances \cite{Jacobson:2008aj}. In our present study we consider these theories to guide ourselves to deal with a cosmological scenario where one can flow from small to large distances under  a Lorentz-violating background. As we shall discuss below, this scenario captures the equations of state of both inflation and dark energy at asymptotic limits, which confirms the aforementioned efforts of producing accelerating Universe in the presence of Lorentz invariance violation.}

In this paper we shall focus on a theoretical approach that deals with  
inflaton field in an early inflationary and late-accelerating Universe. The main point is that a time-like Lorentz-violating background permeating the space is responsible for  
dark energy at large distances (low redshifts). {The reason to work with a time-like Lorentz operator, is that a purely time-like background (or frame), is described by a small subset of Lorentz invariant violating operators which preserves rotational invariance --- see for instance, Passos {\it et al.} \cite{passos} for recent discussion. Besides, as discussed  by Kosteleck\'y and Mewes \cite{kost_09}, the frame of the Cosmic Microwave Background stands as a natural choice for a preferred frame in this context. }

\section{The model}

Let us start with the Lagrangian of a $3+1$-dimensional field theory in a curved background describing an inflaton field  
coupled to Lorentz-violating background
\be\label{act1}
    e^{-1}\LL = \frac{R}{2
    \kappa^2}-\dfrac{1}{2}\Big(g_{\mu\nu}+
    \xi_1 k^1_ {\mu\nu}\Big)\partial^{\mu}\phi\partial^{\nu}\phi
    - V(\phi),
\ee 
where $e=\sqrt{-g}$ and $\kappa^2=8\pi G$. $V(\phi)$ is the inflaton
potential. 
The time-like tensors $k^i_ {\mu\nu}\ ( i=1,2, ...)$ that in principle could couple to several other fields in a more general Lagrangian are such that only $k^i_{00}\neq0$, that is 
\be k^i_ {\mu\nu}=\left(  \begin{array}{cc}
-\beta_i & 0 \\ 
0 & 0
\end{array} \right), 
\ee
where we have defined $k^1_ {\mu\nu}$ as the tensor related to the inflaton. 
The tensor couples the inflaton field to the Lorentz-violating background 
through the coupling $\xi_1>0$.
 It is expected that at short distances 
 $\beta_1\to0$ maintains all the inflationary dynamics depending only on the inflaton field, whereas for  $\beta_1\to-1$, the  Lorentz-violating 
 background becomes more effective at large distances  
 and develops dark energy, as we shall see shortly.

We shall study cosmological issues in this model, so that our {\it  background metric} is the {Friedmann-Robertson-Walker (FRW)} metric of a $flat$ Universe given as
\be
ds^ 2 = g_{\mu\nu}dx^\mu dx^\nu=-dt^ 2+a(t)^2(dx^ 2+dy^ 2+dz^ 2).
\ee
The {\it effective metric} due to presence of the background field is
\be
d\tilde{s}^ 2 = \tilde{g}_{\mu\nu}dx^\mu dx^\nu=\Big( g_{\mu\nu}+\xi_1k_{\mu\nu}\Big)dx^\mu dx^\nu=-\Big(1+\xi_1\beta_1\Big)c^2dt^ 2+a(t)^2(dx^ 2+dy^ 2+dz^ 2),
\ee
where $c$ is the speed of light. The effective velocities for inflaton  
due to the background field are given by 
\be\label{v_i}
v_1=\sqrt{1+\xi_1\beta_1}c. 
\ee
The present choice of Lorenz-violating parameter just change the Lorentz boosts because affects only velocities. 

Thus, in this sense we can define an analogous refractive index felt by the inflaton field as follows:
\be\label{index1}
n\equiv \frac{c}{v_1}=\frac{1}{\sqrt{1+\xi_1\beta_1}}.
\ee
Since the refractive index is in general a wave-length dependent quantity given basically by
\be
n^2=1+\sum_{i=1}^NB_i\frac{\lambda^2_{0i}\lambda^2}{\lambda^2-\lambda^2_{0i}}=1+NB_0\frac{\lambda^2_{0}\lambda^2}{\lambda^2-\lambda^2_{0}},
\ee
{where in the last step we have assumed that all ``molecules'' resonate with same frequency, which means that $B_i=B_0$. This is an analog of the Sellmeier dispersion equation \cite{hecht} which is well-known in optics. In the original formula, $B_i$ are known as Sellmeier coefficients that are determined experimentally. }From the cosmological perspective we make the identification $\frac{\lambda_0}{\lambda}\equiv\frac{a_0}{a(t)}$, for very small variations in time, where $\lambda_0$ and $\lambda$ are the wavelengths  observed now and then, respectively. {Recalling the definition of redshift $z+1=\frac{\lambda_0}{\lambda}$, with $\lambda\leq \lambda_0$, we find
\be\label{index2}
n^2=1-\frac{C_1}{1-(z+1)^2},
\ee
where $C_1=NB_0\lambda^2_{0}$. The extra minus before $C_1$ above is just to keep the analogy with Sellmeier equation normally applied, e.g., to glass for $\lambda\geq\lambda_0$. Now comparing (\ref{index2}) with  (\ref{index1}) we establish a relationship between the Lorentz-violating parameter and the cosmological redshift as follows
\be\label{red-LIV}
\xi_1\beta_1(z)=-\frac{C_1}{z(z+2)+C_1}.
\ee
Notice that with (\ref{red-LIV}) into (\ref{v_i})  the inflaton field has velocity $v(z)\approx0$ for very small redshifts but approaches velocity of light for very large redshifts. } Thus, is expected that in the reheating phase (at UV) the field is expected to develop a radiation-dominated phase, whereas in the current regime (at IR) is expected to be responsible for the dark energy.

\section{From inflationary to dark energy regime}

Let us now study the regime where the inflaton is the dominant species.  The Einstein field equations 
lead to the Friedmann equation $H^2=\frac{8\pi G}{3}\rho_\phi$ whose  inflaton density $\rho_\phi$ is governed by the inflaton field with energy-momentum tensor
\be\label{Tmunu} T_{\mu\nu}=\partial_\mu \phi
\partial_\nu \phi + g_{\mu\nu} \LL_\phi, \qquad T^{\mu}_\nu={\rm diag}\Big({-\rho_\phi, p_\phi, p_\phi, p_\phi}\Big)\,, \ee
{where $\cal{L}_\phi$ is
\be
{\cal L}_\phi=-\dfrac{1}{2}\Big(g_{\mu\nu}+
    \xi_1 k^1_ {\mu\nu}\Big)\partial^{\mu}\phi\partial^{\nu}\phi-V(\phi)\,,
\ee
and $p_{\phi}$ is the pressure component of the energy-momentum tensor due the scalar field. From now on, we are going to work with $c=1$.}
Since we will be interested in homogeneous inflaton configurations $\phi\equiv\phi(t)$,
we can write down the energy density $\rho_\phi$ and pressures $p_i=p_\phi\  (i=1,2,3)$ as follows
\bes\label{prho} \ben
\rho_\phi=\dfrac{1}{2}\Big(1-\xi_1\beta_1\Big)\dot{\phi}^ 2 +V(\phi), \\
p_\phi=\dfrac{1}{2}\Big(1+\xi_1\beta_1\Big)\dot{\phi}^ 2 -V(\phi), \een \ees
where the dot stands for derivative with respect to the temporal coordinate.

The equation of state for the inflaton field can be readily found and is given by
\be\label{eq-state-w}
\omega\equiv\frac{p}{\rho}=\frac{\dfrac{1}{2}\Big(1+\xi_1\beta_1\Big)\dot{\phi}^ 2 -V(\phi)}{\dfrac{1}{2}\Big(1-\xi_1\beta_1\Big)\dot{\phi}^ 2 +V(\phi)}.
\ee
Note that when the potential part dominates one finds the usual $\omega\simeq-1$ and an inflationary regime takes place, where we may constrain $\xi_1\beta_1$ via calculation of the number of $e$-folds --- see below. On the other hand, when the kinetic part {\it at the end of inflation} dominates over the scalar potential part we find the interesting equation of state
\be\label{eq-1.2-inflation}
\omega\simeq\frac{1+\xi_1\beta_1}{1-\xi_1\beta_1},
\ee
that agrees with radiation-dominated regime $\omega\to\frac13$ as $\xi_1\beta_1\to-\frac{1}{2}$ and also matter-dominated regime $\omega=0$ at $\xi_1\beta_1\to-1$. These values are completely in accord with (\ref{red-LIV}) since $\xi_1\beta_1\in(-1,0)$. Furthermore, it is interesting to notice that in general the equation of state (\ref{eq-state-w}) has the following behaviors by power expanding it around $z=0$ (IR) and  $z=\infty$ (UV)
\be\label{eq-st-z}
\omega_0(z)=-\frac{V}{\dot{\phi}^2+V}+{\cal O}(z),\qquad \omega_\infty(z)=\frac{\frac12\dot{\phi}^2-V}{\frac12\dot{\phi}^2+V}+{\cal O}\left(\frac{1}{z^2}\right)
\ee
with $\omega_\infty(z)\in(-1,1)$ and $\omega_0(z)\in(-1,0)$ which means that in the {\it slow-roll} regime, {where $\dot{\phi}^{\,2} \ll V(\phi)$},  the equation of state at inflationary phase $\omega_\infty(z)\simeq-1$ is somehow redshifted to the equation of state of dark energy $\omega_0(z)\simeq-1$.

Let us now focus on such aforementioned regime at inflationary phase. The modified equation of motion for the inflaton field is 
\be
\ddot{\phi}+3H\dot{\phi}+\Big(\dfrac{1}{1+\xi_1\beta_1}\Big)\,V^{\,\prime}=0\,; \qquad V^{\,\prime} = \dfrac{\partial V}{\partial\phi}\,,
\label{mov1}
\ee
whereas the modified Friedmann equation reads
\be
H^2=\dfrac{8\pi G}{3}\Big[\dfrac{1}{2}\Big(1-\xi_1\beta_1\Big)\dot{\phi}^ 2+V(\phi)\Big].
\label{Frid1}
\ee
Note the presence of the Lorentz-violating background  
in both equations which will drive new effects as we shall see below.
In the equation \eqref{mov1}  we may find a slow-roll regime, that is, when the friction term $3H\dot{\phi}$ dominates over the acceleration term $\ddot{\phi}$. This is also accompanied by the condition $\dot{\phi}^ 2\ll V(\phi)$ into (\ref{Frid1}). Thus we find now the following equations 
\be
3H\dot{\phi}+\Big(\dfrac{1}{1+\xi_1\beta_1}\Big)\,V^{\,\prime}\simeq 0,\,\,\,H\equiv \Big(\dfrac{d\ln a}{dt}\Big)\simeq\sqrt{\dfrac{8\pi G}{3}V(\phi)}.
\label{mov2}
\ee
Let us now consider the simplest inflaton potential, the quadratic potential 
\be\label{pot-quad}
V(\phi)=\dfrac12m^2\phi^ 2.
\ee
For the slow-roll condition the equation \eqref{Frid1} simplifies to
\be
H=\sqrt{\dfrac{4\pi G}{3}}\ m\phi,
\label{mov3}
\ee 
and the equation \eqref{mov2} leads to
\be 
\dot{\phi}=-\dfrac{m}{\sqrt{12\pi G}(1+\xi_1\beta_1)},
\ee
that is
\be \label{sol}
\phi(t)=\phi_{0}-\dfrac{m}{\sqrt{12\pi G}(1+\xi_1\beta_1)}t.
\ee
Finally using \eqref{mov3} we obtain the scale factor
\be
a(t)=a_ {0}\exp{\left[\sqrt{\dfrac{4\pi G}{3}}m\phi_{0}t-\dfrac{m^ 2}{6(1+\xi_1\beta_1)}t^ 2\right]}.
\ee

Now to understand the effect of the Lorentz-violating background on the inflationary phase of the Universe we make use of the $e$-folds number defined as 
\be
N_{e}=\int_{t_{i}}^ {t_{f}}Hdt,
\label{folds}
\ee
that making use of (\ref{mov2}) we find 
\be
N_{e}=\int_{t_{i}}^ {t_{f}}H\,dt= \int_{t_{i}}^ {t_{f}}H\,\frac{d\,\phi}{d\,\phi}\,dt=\int_{\phi(t_{i})}^ {\phi(t_{f})}\frac{H}{\dot{\phi}}\,d\phi\,
\ee
so
\be
N_{e}=-\frac{3}{1+\xi_1\,\beta_1}\,\int_{\phi(t_{i})}^ {\phi(t_{f})}\,\frac{H^{\,2}}{V^{\,\prime}}\,d\phi=-\frac{8\,\pi\,G}{1+\xi_1\beta_1}\,\int_{\phi(t_{i})}^ {\phi(t_{f})}\,\frac{V}{V^{\,\prime}}\,d\phi\,.
\ee
For the inflaton potential defined in (\ref{pot-quad}) we find the modified $e$-fold number
\be
N_{e}=-\dfrac{4\pi G}{1+\xi_1\beta_1}\int_{\phi(t_{i})}^ {\phi(t_{f})}\phi \, d\phi=\dfrac{2\pi G}{1+\xi_1\beta_1}\Big(\phi(t_{i})^ 2-\phi(t_{f})^ 2\Big).
\ee
Now assuming  $\phi(t_{f})\approx 0$ we obtain
\be \label{nefold}
N_{e}=\frac{2\pi G}{1+\xi_1\beta_1}\phi(t_{i})^ 2,
\ee
that implies
\be
\xi_1\beta_1=\dfrac{2\pi G\phi(t_{i})^ 2}{N_{e}}-1.
\label{plank}
\ee
For $N_{e}\simeq60$, $\phi(t_{i})^ 2\simeq 4m^ {2}_{pl}$ and $G\simeq \dfrac{1}{m^ {2}_{pl}}$ we find  for \eqref{plank}
\be
\beta_1\simeq\left(\dfrac {8\pi}{60}-1\right)\xi_1^{-1}\simeq -\frac{0.5811}{\xi_1}, \qquad \xi_1>0.
\ee
Notice this agrees with the end of inflation (and also with the beginning of the radiation) regime described by Eq.~(\ref{eq-1.2-inflation}) where Lorentz-violating background field assumes the value $\xi_1\beta_1=-1/2$. 

{
Besides the approaches above, there are several interesting cosmological properties which may be derived from the slow-roll parameters, as well as, from the power spectrum perturbations. Let us first observe the consequences of this inflationary model for the two first slow-roll parameters, whose explicit forms in the slow-roll regime  are \cite{ellis}
\be
\epsilon=\frac{1}{16 \pi  G}\,\left(\frac{V^{\,\prime}}{V}\right)^2\,; \qquad \eta = \frac{1}{8 \,\pi\,  G}\,\frac{V^{\,\prime\,\prime}}{V}\,.
\ee
The previous equations yield to
\be \label{slowroll}
\epsilon=\eta=\frac{1}{4\,\pi\,G\,\phi^{\,2}}\,,
\ee
which is consistent with a chaotic inflation driven by a quadratic potential. As in the standard inflationary scenario, the strength of the tensor perturbations is directly related with the magnitude of the energy density. It is well-known that the power spectrum for scalar perturbation has the form \cite{ellis}
\be
P_{\zeta}=\frac{H^{\,4}}{4\,\pi^{\,2}\,\dot{\phi}^{\,2}}\,,
\ee
when we deal with one scalar field Lagrangians. It is important to point that all quantities here are determined at the horizon crossing \cite{ellis, mukhanov}. Such a parameter enable us to compute the so-called  scalar spectral index, which is given by \cite{ellis}
\be \label{ssp0}
n_s-1\equiv \frac{1}{H\,P_{\,\zeta}}\,\frac{d\,P_{\zeta}}{d\,t}\,,
\ee
and this parameter is remarkably important as a test for cosmological models, since it is directly measured in the CMB. An equivalent way to represent $n_s$ is through the expression
\be
n_s=-6\,\epsilon +2\,\eta = -4\,\epsilon
\ee
where we use the fact that $\epsilon=\eta$ due the slow-roll approximation.  The previous equation together with $(\ref{nefold})$, and $(\ref{slowroll})$ allows us to write $n_s$ as
\be \label{ssp}
n_s-1=-\frac{2}{1+\xi_1\,\beta_1\,N_e}\approx -\frac{2}{N_e}+\delta_{n_s}\,; \qquad \delta_{n_s}=\frac{2}{N_e}\,\xi_1\,\beta_1\,.
\ee
where we consider $\phi(t) = \phi(t_i)$ in Eq. $(\ref{slowroll})$ at the horizon crossing, moreover we also expanded the right-hand side around small values of $\xi_1\beta_1$.

Recently, Planck collaboration \cite{planck15} strongly constrained that the scalar spectral index has the value $n_s=0.9655\,\pm\,0.0062$, therefore, we are able to use $(\ref{ssp})$ to establish a bound for the product $\xi_1\beta_1$ related with the CMB data. This bound is strongest than the one derived from $(\ref{plank})$, once we do not need to impose any initial value for $\phi(t_i)$.  Following the procedure adopted in \cite{manoel}, we assume the equalities 
\be
n_s=1-\frac{2}{N_e}=0.9655\,; \qquad \delta_{n_s}=\frac{2}{N_e}\,\xi_1\,\beta_1 \sim -10^{\,-3}\,,
\ee
where we associated the Lorentz-breaking parameter with the order of the experimental error for $n_s$. Therefore, we find that $N_e \sim 57.97$ and $\xi_1\,\beta_1\sim - 10^{\,-2}$\,. From  $(\ref{mov3})$, $(\ref{sol})$, and $(\ref{ssp0})$, we derive $n_s$ as a function of time, yielding to
\be
n_s=1-\frac{36 (\beta_1\,\xi_1+1)}{\left(\sqrt{3}\, m\, t-6\, \sqrt{\pi\,G} \phi_0 (\beta_1\,\xi_1+1)\right)^2}\,.
\ee
This last result is going to be useful to determine a relation between the scalar spectral index with the denominated tensor-scalar ratio, whose definition is
\be
r=\frac{P_T}{P_{\,\zeta}}\,; \qquad P_T=64\,\pi\,G\,\left(\frac{H}{2\,\pi}\right)^{\,2}\,.
\ee
Thus, taking Eqs. $(\ref{mov3})$, and $(\ref{sol})$,  the explicit time dependence of $r$ is
\be
r=\frac{144}{\left(\sqrt{3}\, m\, t-6\, \sqrt{\pi\,G} \phi_0 (\beta_1\,\xi_1+1)\right)^2}\,,
\ee
therefore, by combining $n_s$ with $r$ we are able to find the relation
\be
r=\frac{4 (1-n_s)}{\beta_1\, \xi_1+1}\,,
\ee
unveiling that the Lorentz-breaking parameter changes the standard dependence between $n_s$ and $r$ (see Eq. (2.7) at \cite{ellis} for comparison). Furthermore, from $(\ref{ssp})$ we determine that
\be
r=\frac{1}{N_e}\,\frac{8}{(1+\xi_1\,\beta_1)^{\,3}} \approx \frac{8}{N_e}\,\left(1-3\,\beta_1\,\xi_1\right)\,,
\ee
then, using the previous constraints to $N_e$ and to $\xi_1\,\beta_1$, we yield to $r \sim 10^{\,-1}$, which is compatible with the  Planck collaboration measurements for cosmological parameters \cite{planck15}.
}

\section{Conclusions}

We present a new model to address the dark energy problem  by changing the kinetic part of the inflaton field with a Lorentz-violating time-like background that plays the role of a medium that affects the inflaton velocity, {as one can see in Eq. $(\ref{v_i})$. Such a behavior is responsible to change how long the inflationary process can last}. By identifying an analogous refractive index, {see Eq. $(\ref{index1})$, we were able to relate this background with the cosmological redshift. Such a relation lead us to conclude that a time-like Lorentz-violating background can be responsible for 
dark energy at low energy (low redshifts). Furthermore, this is in accord with the aforementioned fact that one has been shown that violation of the Lorentz invariance induces Lagrangians that can drive the present acceleration of the universe.} More interesting, the combined slow-roll and dominated-Lorentz-violating background regimes develop an equation of state for the dark energy which approaches the equation of state of the cosmological constant, as pointed in Eq. $(\ref{eq-st-z})$. {This feature is remarkably consistent with the expansion phase that our Universe has been passing through. Another interesting result of our investigation was the determination of the scalar spectral index $(n_s)$ and of the tensor-scalar ratio $(r)$, for this cosmological scenario. Note that the Lorentz-breaking parameter shifted the relation between both cosmological parameters. Moreover, the bound for the product $\xi_1\,\beta_1$ derived from $n_s$, resulted in a value for tensor-scalar ratio which agrees with Planck's data for cosmological parameters. Such a test strength the potential of our work, and further studies should be addressed elsewhere.}


{\acknowledgments} 

We would like to thank CNPq for partial financial support, and also the anonymous referees for their fruitful comments which enhanced the potential of this work.

\end{document}